\documentclass{ws-procs975x65}

\usepackage{graphicx}
\usepackage{amsmath}
\usepackage{amssymb}
\usepackage{subfigure}

\begin{document}

\title{Super Massive Black Holes and the Origin of High-Velocity Stars}
\author{Roberto Capuzzo-Dolcetta$^*$ and Giacomo Fragione$^{**}$}

\address{Dep. of Physics, Sapienza, Univ. of Roma\\
P.le A. Moro 2, Roma, Italy\\
$^*$E-mail: roberto.capuzzodolcetta@uniroma1.it\\
$^{**}$E-mail: giacomo.fragione90@gmail.it}

\begin{abstract}
The origin of high velocity stars observed in the halo of our Galaxy is still unclear. In this work we test the hypothesis, raised by results of recent high precision $N$-body simulations, of strong acceleration of stars belonging to a massive globular cluster orbitally decayed in the central region of the host galaxy where it suffers of a close interaction with a super massive black hole, which, for these test cases, we assumed $10^8$ M$_\odot$ in mass.

\end{abstract}


\bodymatter

\section{Introduction}

Hypervelocity stars (HVS) are stars escaping the host galaxy. Hills \cite{hil88} was the first to predict theoretically their existence as a consequence of interactions with a massive Black Hole (BH) in the Galactic Centre \cite{hil88}, while Brown et al. serendipitously discovered the first HVS in the outer stellar halo of the Galaxy, a B-type star moving over twice the Galactic escape velocity \cite{brw05}. The most recent HVS Survey is the Multiple Mirror Telescope Survey, which revealed $21$ HVSs at distances between $50$ and $120$ kpc \cite{brw14}.

Hills' mechanism involves the tidal breakup of a binary passing close to a massive BH, which could lead also to a population of stars orbiting in the inner regions of the Galaxy around the central BH, the so-called S stars \cite{per07}. Since the Hills' prediction, many other mechanisms have been proposed to explain the production of HVSs, which involve different astrophysical frameworks and phenomena \cite{yut03,tuf09}. The study of the characteristics of these stars would help to infer information on both the small and large scales of the Galaxy, i.e. the region near massive BHs as well the shape of the Galaxy and Dark Matter gravitational potential \cite{gne05}.

The aim of the present work is to investigate another mechanism of production of HVS, which involve a Globular Cluster (GC) that during its orbit has the chance to pass close to a super massive black hole (SMBH) in the center of its host galaxy.

\section{Close Globular Cluster-Super Massive Black Hole Interactions}

From direct $N$-body simulations of a GC passing close to an SMBH \citep{AS15}, there is evidence that some GC stars are ejected in sort of jets. Therefore, in order to understand the underlying physical mechanism leading to such ejections, we performed 3-body scattering experiments involving an SMBH, a GC and a star. In our simulations the BH  is initially set in the origin of the reference frame, while the GC (considered as a point mass) follows an elliptical orbit around it within the SMBH influence radius. This assumption is justified by that the GC has had the time to shirnk significantly its orbit by the dynamical friction braking exterted by the stars of the galaxy. We selected a circular orbit of radius $r_c=10$ pc as reference, and sampled a set of GC orbits of same energy, but different eccentricity ($e$), just varying the ratio $0\leq L/L_c\leq1$, where $L$ and $L_c$ are, respectively, the generic orbit angular momentum ($0<e\leq 1$) and that of the circular orbit ($e=0$). Note that $e=\left[1-(L/L_c)^2\right]^{1/2}$.

In the frame of a restricted 3-body problem, the zero velocity (Hill's) surfaces enclose the two finite-mass (SMBH and GC) bodies, dividing the space in a region of influence of the GC and in a region of influence of the BH. A meaningful study refers to the fate of stars moving around the GC with orbits initially lying all within the GC influence radius. Therefore, we consider stars on (initially) circular orbits within this sphere, selecting a set of initial positions at evenly spaced angles along the orbital circumference. While, in our simulations, the BH mass and the test star mass are fixed to $M_{BH}=10^8$ M$_{\odot}$ and to $m_{*}=1$ M$_{\odot}$, respectively, we varied the GC mass choosing $M_{GC}=10^4$, $10^5$ and  $10^6$ M$_{\odot}$.

\def\figsubcap#1{\par\noindent\centering\footnotesize(#1)}
\begin{figure}[h]
\begin{center}
\parbox{2.1in}{\includegraphics[width=2in]{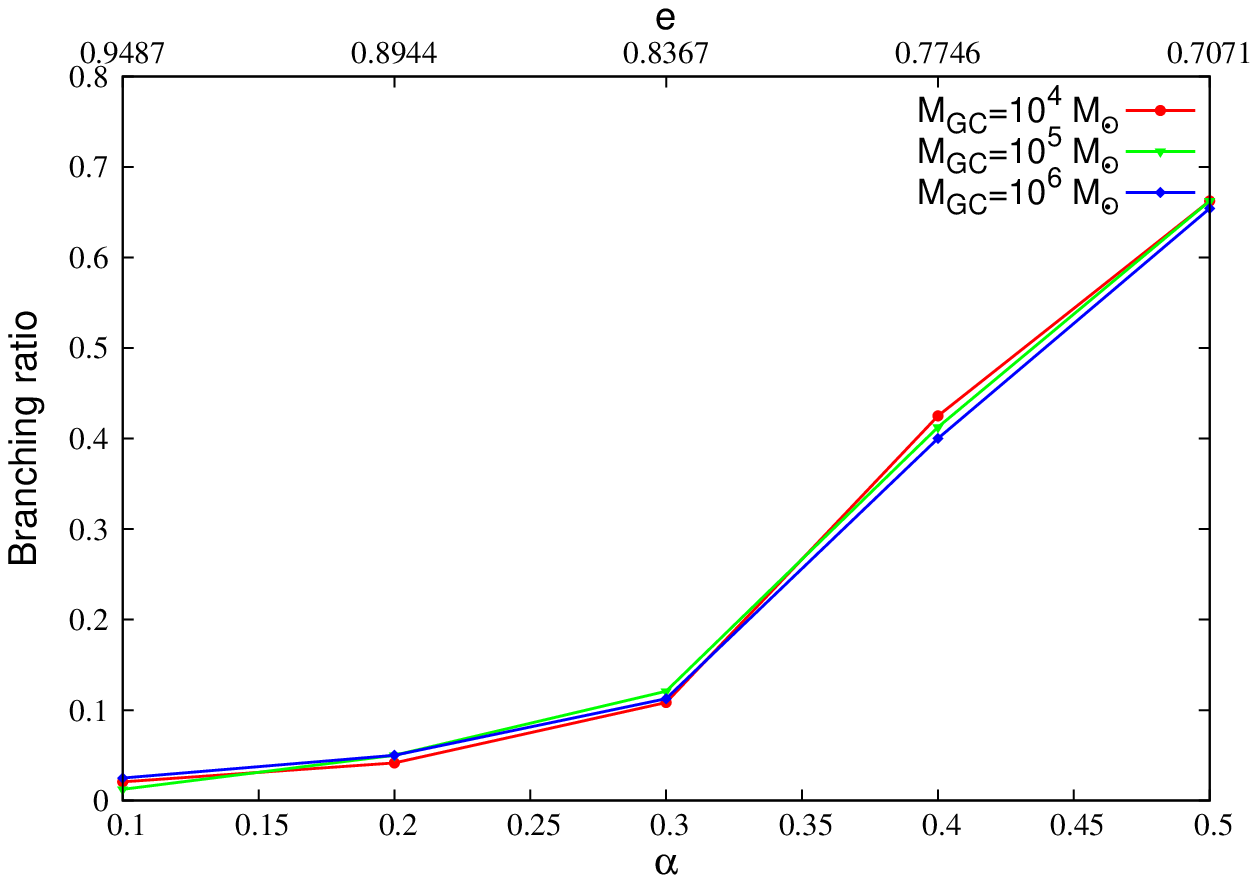}\figsubcap{a}}
\hspace*{4pt}
\parbox{2.1in}{\includegraphics[width=2in]{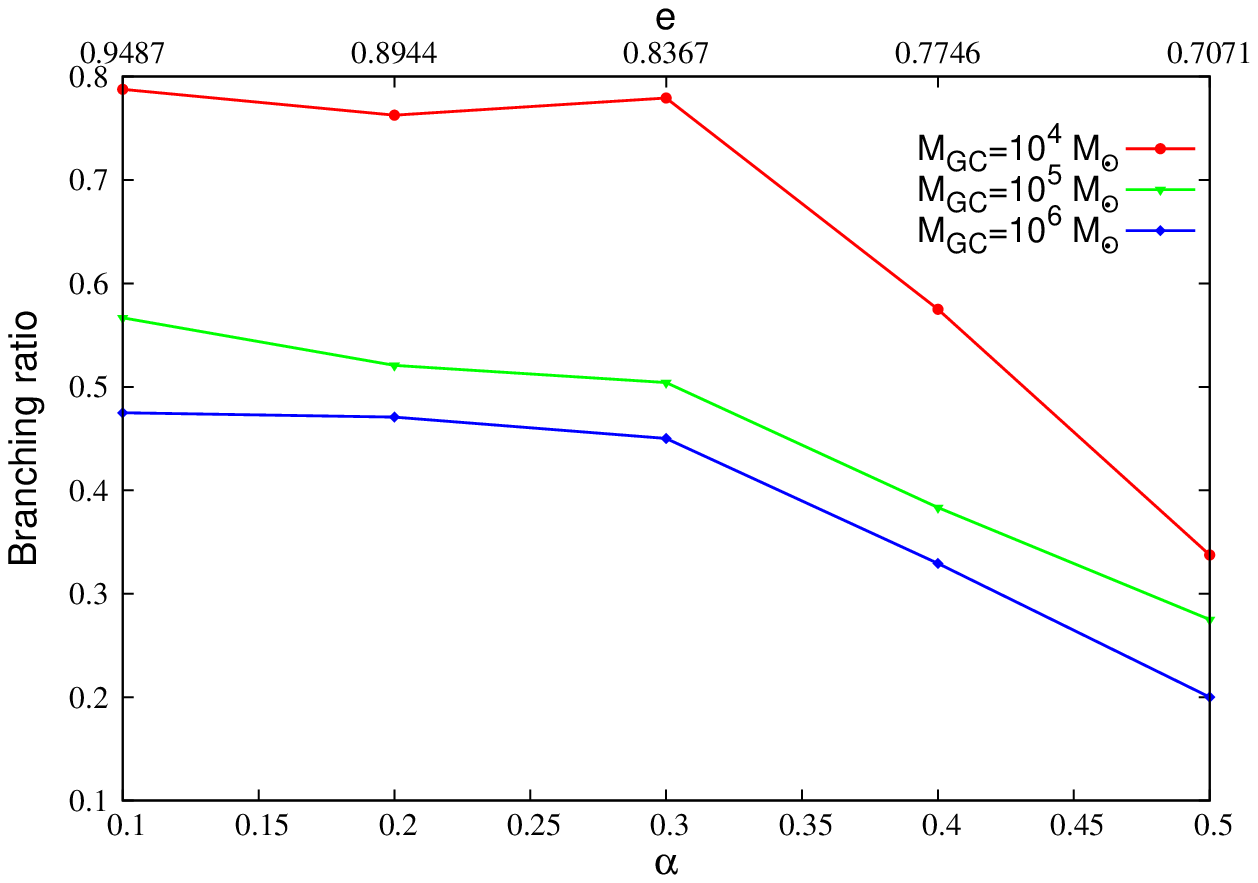}\figsubcap{b}}
\parbox{2.1in}{\includegraphics[width=2in]{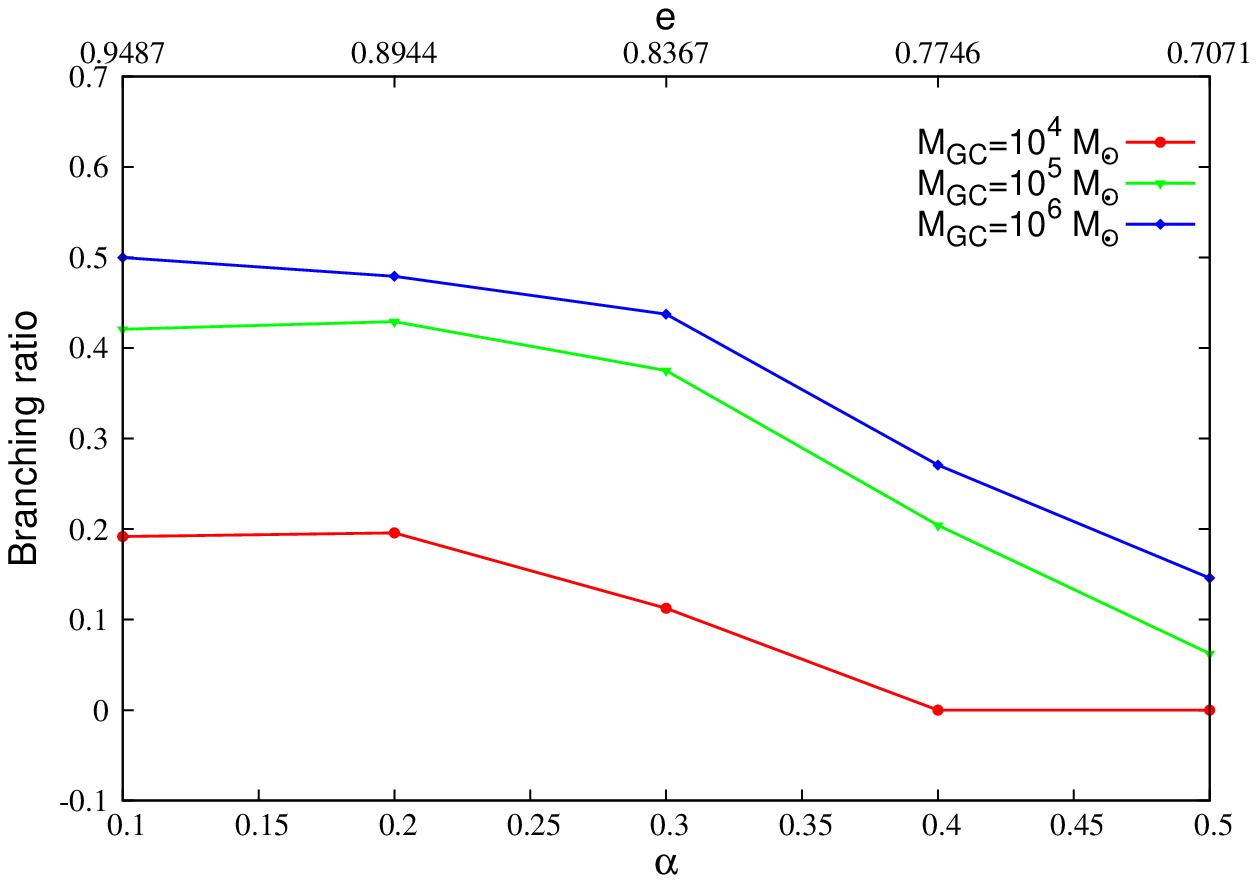}\figsubcap{c}}
\caption{Branching ratios of GC stars (a), S stars (b) and ejected stars (c), after GC-SMBH scattering, for different GC masses and orbits, parametrized by $\alpha=(L/L_c)^2$.}
\label{fig:bratios}
\end{center}
\end{figure}

\def\figsubcap#1{\par\noindent\centering\footnotesize(#1)}
\begin{figure}[h]
\begin{center}
\parbox{2.1in}{\includegraphics[width=2in]{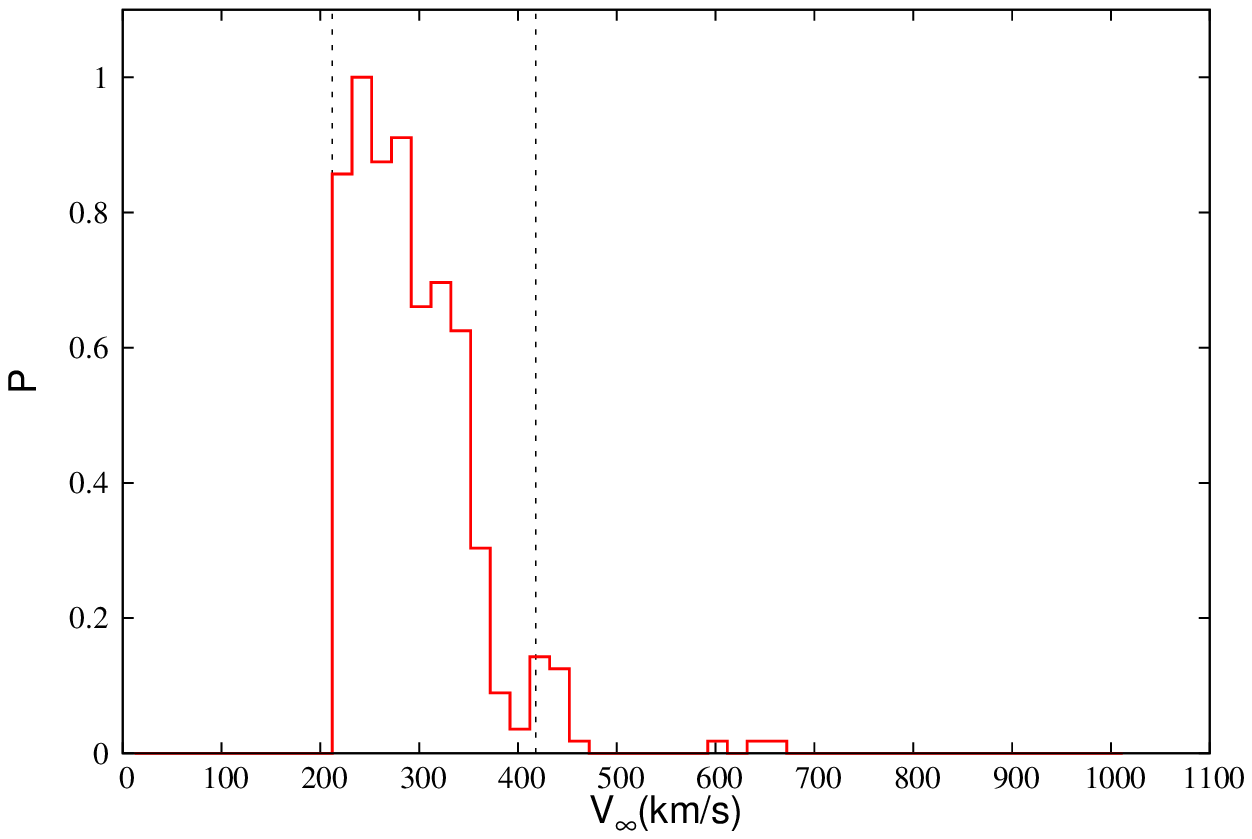}\figsubcap{a}}
\hspace*{4pt}
\parbox{2.1in}{\includegraphics[width=2in]{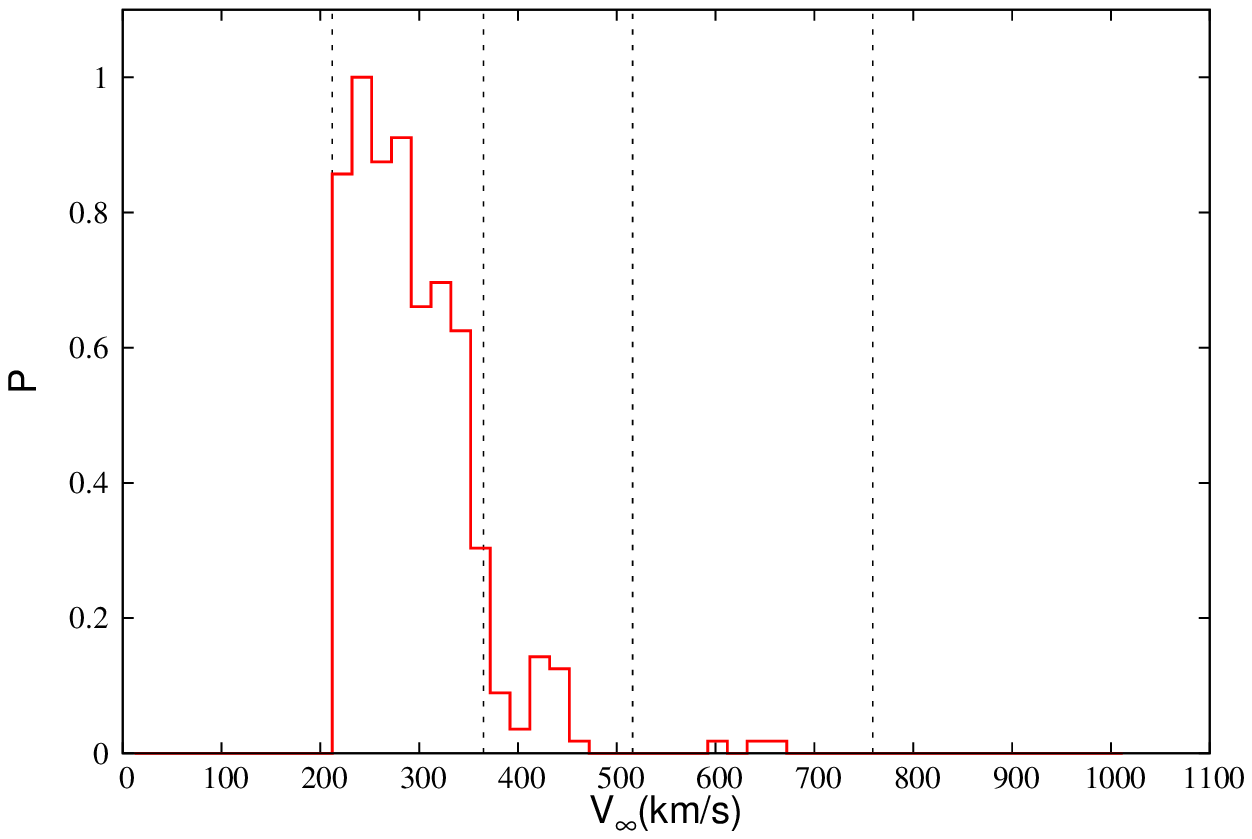}\figsubcap{b}}
\parbox{2.1in}{\includegraphics[width=2in]{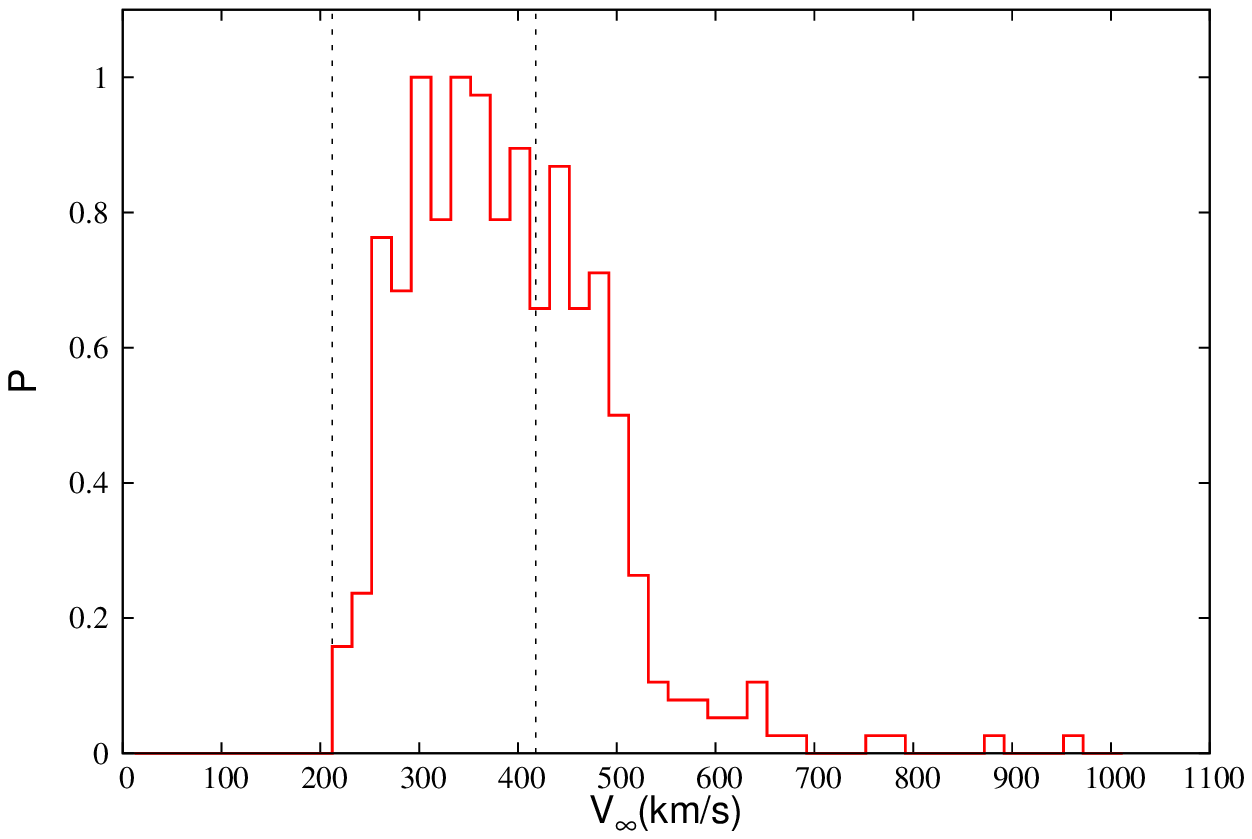}\figsubcap{c}}
\hspace*{4pt}
\parbox{2.1in}{\includegraphics[width=2in]{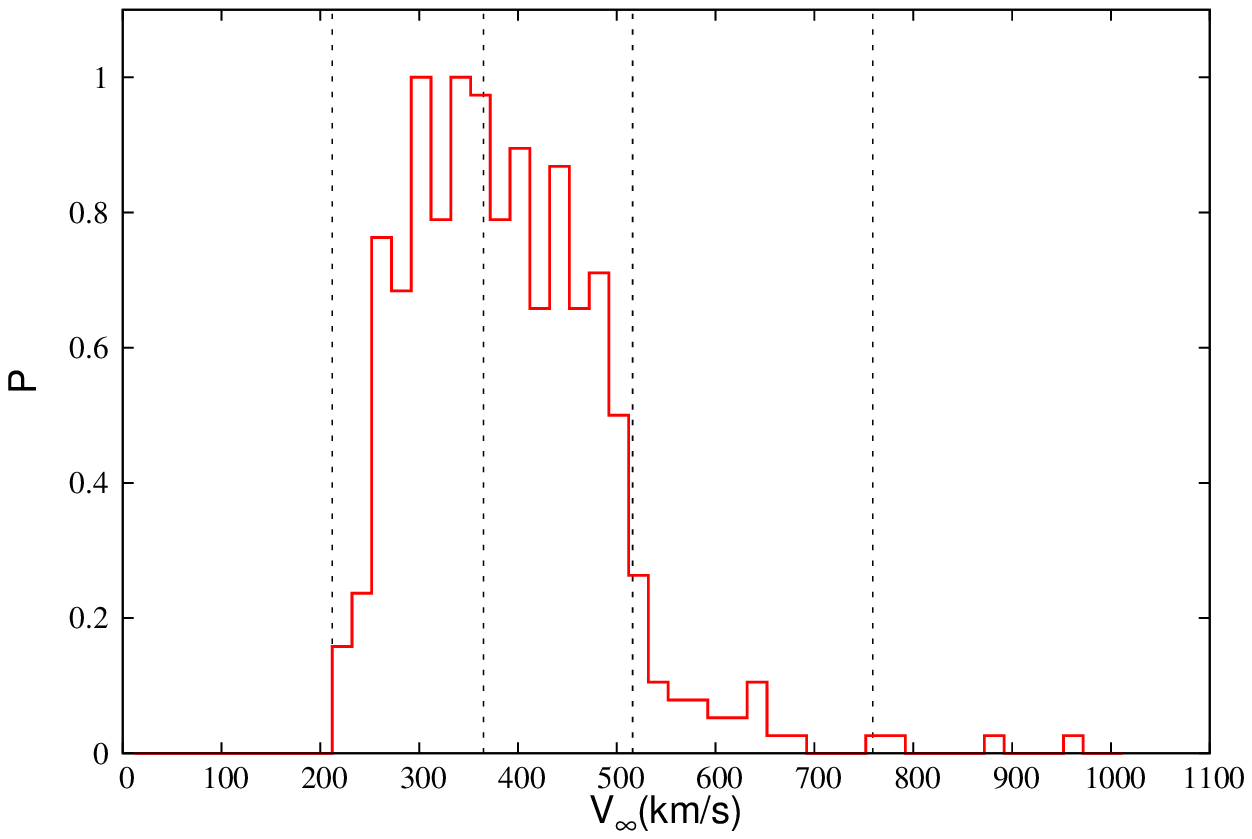}\figsubcap{d}}
\caption{Velocity distribution of escaping stars for $M_{GC}=10^5$ M$_\odot$ (a,b) and $10^6$ M$_\odot$ (c,d) and all the orbits, both for a $M_{tot}=7.81 \times 10^{10}$ M$_{\odot}$ elliptical galaxy \cite{mar03} (left column) and a $M_{tot}=6.60 \times 10^{11}$ M$_{\odot}$ spiral \cite{fuj09} (right column). The leftmost dashed line in all the panels indicates the escape velocity from the SMBH ($212\ km\ s^{-1}$). In the left column panels, the other vertical line refers to the escape velocity ($418\ km\ s^{-1}$) from the BH +  bulge-halo system, while, in the right panels, the other vertical lines indicate the escape velocity respect to the BH + bulge ($365\ km\ s^{-1}$), BH + bulge + disk ($516\ km\ s^{-1}$), BH + bulge + disk + dark halo ($759\ km\ s^{-1}$), respectively.}
\label{fig:velocity}
\end{center}
\end{figure}

Given the above set of initial parameters, we integrated the system of the differential equations of the 3-bodies motion using the fully regularized algorithm of Mikkola and Aarseth \cite{mik01}. The need of a regularized algorithms is due to the enormous range of variation of the masses involved, which span the $1 \div 10^8$ range. 

The test star orbiting the GC has three possible fates after the GC-SMBH encounter: (a) it remains bound to the GC on an orbit perturbed respect the original one; (b) it becomes a {\it high velocity} star, either bound or unbound; (c) it becomes a {\it S star}, i.e. it is captured by the BH gravitational field and starts revolving around it. 
The branching ratios of these 3 different scattering results are plotted in Fig.\ref{fig:bratios}.

If the star escaping from the GC passes through the first Lagrangian libration point, L1, its fate is the capture by the BH, while when crossing L2 it will escape the whole (GC + BH) system. The first channel is favoured by smaller GC to BH mass ratios, since the BH potential is stronger and is able to capture a larger number of GC stars making them pass through L1. At the same time, when the GC mass is not large, the GC gravitational potential is not intense enough to give the star, escaping it, a velocity sufficient to escape the whole GC+BH system. Therefore the branching ratio for S stars production is higher for lower GC masses, while that of ejected stars increases for higher GC masses.

To evaluate whether stars formerly belonging to the GC and emitted at a high velocity are actually bound or unbound to the host galaxy, we need an assumption on the galactic field. We assumed two different models for the host galaxy, one as an elliptical and one as a spiral galaxy. The elliptical galaxy potential is represented by a two-component model \cite{mar03} (SMBH+spherical bulge-halo), with $M_{tot}=7.81 \times 10^{10}$ M$_{\odot}$, while the spiral galaxy is represented as a four-component model \cite{fuj09} (SMBH+spherical bulge+axisymmetric disk+spherical halo), with $M_{tot}=6.60 \times 10^{11}$ M$_{\odot}$. The results are plotted in Fig.\ref{fig:velocity}, which shows that some of the ejected stars are HVS, i.e. they are unbound respect to the galactic potential. The fraction of HVS depends of course on both the shape of the galactic potential and on the total mass of the host galaxy.

\section{Conclusions}
In this paper we deepened what has been recently found by Ref. \refcite{AS15}, i.e. that the close passage of a massive globular cluster near to a massive black hole can be source of ejection of stars from the cluster, which are accelerated to high speed. The underlying mechanism is likely a 3-body interaction, where the~\lq bodies\rq\ are the super massive black hole ($10^8$ M$_\odot$), the globular cluster ($10^4$, $10^5$ and $10^6$ M$_\odot$) and the test star (1 M$_\odot$) belonging to the globular cluster. 
We adopted a high mass for the BH with the scope of identify at better the underlying physical mechanism. Main preliminary results are:
\begin{itemize}
\item 
the efficiency of the star acceleration process is almost linear in $M_{GC}$;
\item 
given a massive globular cluster (composed by $10^6$ identical 1 M$_\odot$ stars), it releases, in a single close passage around the super massive black hole, about $10^4$ stars;
\item 
in a very close GC-BH encounter ($\alpha=0.1$, $M_{GC}=10^6$ M$_\odot$) the fractions of stars which remain bound, become S stars, escape from the cluster are $\sim 5\%$, $\sim 45\%$, $\sim 50\%$, respectively;
\item
the fraction of stars which escape from the whole galaxy is $\sim 18\%$ ($\sim 0.5\%$) for an $M_{tot}=7.81 \times 10^{10}$ M$_{\odot}$ elliptical ($M_{tot}=6.60 \times 10^{11}$ M$_{\odot}$ spiral) galaxy.
\end{itemize}

\end{document}